\documentclass[aps, pra, reprint]{revtex4-2}
\usepackage{graphicx} 
\usepackage{amsmath, physics}
\usepackage{amssymb}
\usepackage{amsfonts}
\usepackage[unicode]{hyperref}
\hypersetup{
   unicode=true,          %
   plainpages=false,
   colorlinks=true,       
   citecolor=blue,        
   urlcolor=blue
}
\usepackage[caption=false,position=top,singlelinecheck=off,justification=raggedright]{subfig}
\usepackage{color}

\begin{document}
\title{Apparent delay of the Kibble-Zurek mechanism in quenched open systems}

\author{Roy D. Jara Jr.}
\email{rjara@nip.upd.edu.ph}
\affiliation{National Institute of Physics, University of the Philippines, Diliman, Quezon City 1101, Philippines}

\author{Jayson G. Cosme}
\email{jcosme@nip.upd.edu.ph}
\affiliation{National Institute of Physics, University of the Philippines, Diliman, Quezon City 1101, Philippines}

\date{\today}

\begin{abstract}

We report an intermediate regime in the quench time, $\tau_{q}$, separating the usual validity of the Kibble-Zurek mechanism (KZM) and its breakdown for rapid quenches in open systems under finite quench protocols. It manifests in the power-law scaling of the transition time with $\tau_{q}$ as the system appears to enter the adiabatic regime, even though the ramp is already terminated and the final quench value is held constant. This intermediate regime, which we dub the delayed KZM, emerges due to the dissipation, preventing the system from freezing in the impulse regime. This results in a large delay between the actual time the system undergoes a phase transition and the time inferred from a threshold-based criterion for the order parameter, as done in most experiments. We demonstrate using the open Dicke model and its one-dimensional lattice version that this phenomenon is a generic feature of open systems that can be mapped onto an effective coupled oscillator model. We also show that the phenomenon becomes more prominent near criticality, and its effects on the transition time measurement can be further exacerbated by large threshold values for an order parameter. Due to this, we propose an alternative method for threshold-based criterion which uses the spatiotemporal information, such as the system's defect number, for identifying the transition time.

\end{abstract}

\maketitle

\section{Introduction}

Initially formulated to describe the evolution of topological defects in the early universe \cite{kibble_topology_1976, kibble_implications_1980, zurek_cosmological_1985}, the Kibble-Zurek Mechanism (KZM) has been successful in describing the dependence of the defect number and duration of a continuous phase transition on the quench timescale, $\tau_{q}$ \cite{del_campo_universality_2014}. In particular, the theory has been tested in multiple platforms, ranging from atomic Bose-Einstein condensates \cite{ye_universal_2018, liu_kibble-zurek_2020, navon_critical_2015, clark_universal_2016, nagy_self-organization_2008, shimizu_dynamics_2018, dziarmaga_tensor_2023, anquez_quantum_2016, yukalov_realization_2015}, spin systems \cite{schmitt_quantum_2022, li_probing_2023, du_kibblezurek_2023, mayo_distribution_2021}, Rydberg atom setups \cite{keesling_quantum_2019, chepiga_kibble-zurek_2021}, and trapped-ion systems \cite{cui_experimental_2016, ulm_observation_2013}. It has also been tested in dissipative quantum systems \cite{rossini_dynamic_2020,  hedvall_dynamics_2017, zamora_kibble-zurek_2020, puebla_universal_2020, griffin_scaling_2012,  bacsi_kibblezurek_2023, mayo_distribution_2021, klinder_dynamical_2015}, and has recently been extended to include generic nonequilibrium systems \cite{reichhardt_kibble-zurek_2022, maegochi_kibble-zurek_2022, yang_universal_2023}.

Under the standard KZM, a generic closed system with a continuous phase transition has a diverging relaxation time, $\tau$, and correlation length, $\xi$, as it approaches its critical point, $\lambda_{c}$. In particular, one expects $\tau$ and $\xi$ to scale as $\tau \propto |\varepsilon|^{-vz}$ and $\xi \propto |\varepsilon|^{-v}$ \cite{del_campo_universality_2014}, respectively, where $\varepsilon = \left( \lambda - \lambda_{c} \right) / \lambda_{c}$ is the reduced distance of the control parameter, $\lambda$, from the critical point, while $v$ and $z$ are the static and dynamic critical exponents, respectively. It is then expected that if the system is linearly quenched via a ramp protocol, $\varepsilon = t / \tau_{q}$, the system will become frozen near $\lambda_{c}$ due to $\tau$ diverging. This motivates the introduction of the adiabatic-impulse (AI) approximation, where the system's dynamics are classified into two regimes \cite{del_campo_universality_2014}. Far from $\lambda_{c}$, the system is in an adiabatic regime, in which its macroscopic quantities adiabatically follow the quench. Near $\lambda_{c}$, the system enters the impulse regime, wherein all relevant observables remain frozen even after passing $\lambda_{c}$. It only reenters the adiabatic regime and transitions to a new phase after some finite time referred to as the freeze-out time, $\hat{t}$, has passed \cite{del_campo_universality_2014}. This occurs after the system reaches the AI crossover point, $\varepsilon(\hat{t})$, setting $\hat{t} \sim \tau\left( \varepsilon(\hat{t}) \right)$ \cite{del_campo_universality_2014}. The KZM predicts that, due to the scaling of $\tau$, $\hat{t}$ and $\varepsilon(\hat{t})$ must follow the scaling laws \cite{del_campo_universality_2014} 
\begin{equation}
\label{eq: kzm_predicted_scaling}
 \hat{t} \propto \tau_{q}^{\frac{vz}{1 + vz}}, \quad \varepsilon(\hat{t}) \propto \tau_{q}^{- \frac{1}{1 + vz}}.
\end{equation}

While the standard KZM has been successful in explaining the dynamics of continuously quenched systems, studies on systems with finite quenches have shown that the mechanism breaks down if the quench terminates quickly at a certain value, $\varepsilon_{f}$ \cite{wang_non-equilibrium_2022, gomez-ruiz_role_2022, gomez-ruiz_full_2020, kou_varying_2023, goo_defect_2021, chesler_defect_2015, del_campo_structural_2010, gomez-ruiz_universal_2019,zeng_universal_2023}. In particular, $\hat{t}$ and $\varepsilon(\hat{t})$ saturate at a finite value as $\tau_{q} \rightarrow 0$, with $\varepsilon(\hat{t}) = \varepsilon_{f}$, and thus $\hat{t} \sim \tau(\varepsilon_{f})$ \cite{zeng_universal_2023}. In Ref.~\cite{zeng_universal_2023}, this breakdown of the KZM is predicted to occur at some critical quench time $\tau_{q, c} = \hat{t}_{\mathrm{fast}} / \varepsilon_{f}$, where $\hat{t}_{\mathrm{fast}}$ is the saturation value of $\hat{t}$ in the sudden quench limit, $\tau_{q} \rightarrow 0$.

Measuring the exact value of $\hat{t}$ and $\varepsilon(\hat{t})$ is a nontrivial task unlike defect counting due to the limitations in detecting the exact time a system reenters the adiabatic regime. As such, it is common to employ a threshold criterion and measure instead the transition time, $\hat{t}_{\mathrm{th}}$, which is the time it takes for an order parameter to reach a given threshold after passing $\lambda_{c}$. The crossover point at the transition time, $\varepsilon(\hat{t}_{\mathrm{th}})$, is similarly defined. For a sufficiently small threshold value, it is assumed that $\hat{t}_{\mathrm{th}}$ is a good approximation for $\hat{t}$. While this method is successful in showing the power-law scaling of $\hat{t}_{\mathrm{th}}$ and $\varepsilon(\hat{t}_{\mathrm{th}})$ as a function of $\tau_{q}$ \cite{ye_universal_2018, anquez_quantum_2016, chesler_defect_2015, shimizu_dynamics_2018, wang_non-equilibrium_2022, klinder_dynamical_2015, baumann_dicke_2010}, it remains unclear whether the inherent deviation between $\hat{t}$ and $\hat{t}_{\mathrm{th}}$ does not lead to any significant effects on the scaling of the KZM quantities for any generic quench protocols.

\begin{figure}
    \centering
    \includegraphics[scale = 0.47]{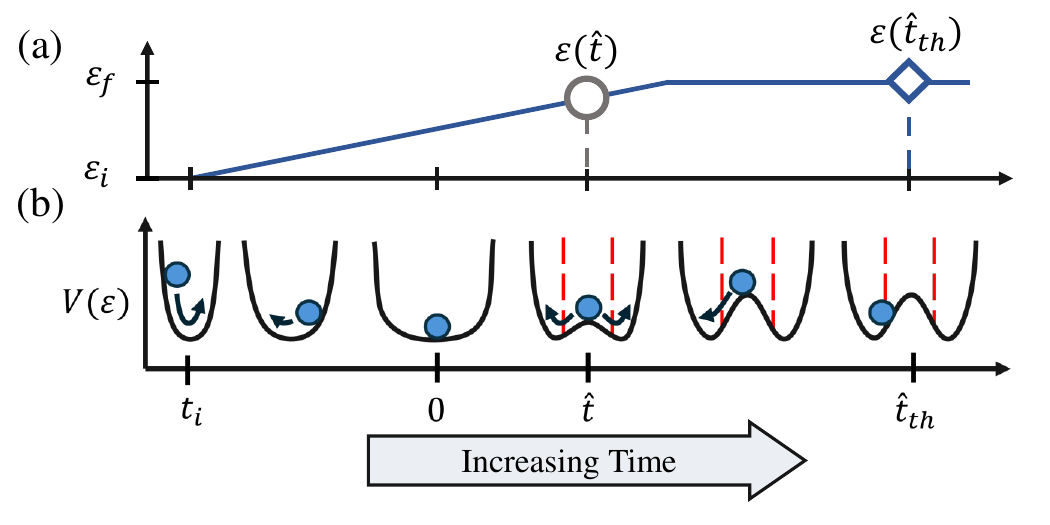}
    \caption{ Sketch of a generic open system undergoing a continuous phase transition. (a) A finite ramp protocol is applied on a system initialized in the normal phase. (b) The quench modifies the effective potential experienced by the system represented by the ball. This pushes the system to enter a symmetry-broken phase at $\hat{t}$ after the ramp passes the critical point $\varepsilon(t = 0) = 0$. The transition, however, is only detected after the system's order parameter reaches the set threshold value, marked by vertical dashed lines, at $\hat{t}_{\mathrm{th}}$.  }
    \label{fig: sketch_result}
\end{figure}

In this paper, we report an intermediate regime separating the breakdown and validity of the KZM appearing in open systems under a finite quench protocol depicted in Fig.~\ref{fig: sketch_result}(a). In this regime, the transition time follows the power-law scaling predicted by the KZM even though the system appears to relax after the quench has terminated, as illustrated in Fig.~\ref{fig: sketch_result}(b). As we will show later, this regime manifests precisely due to the dissipation exacerbating the deviation between the freeze-out time and the transition time, leading to a delay in the detection of the phase transition, as schematically represented in Fig.~\ref{fig: sketch_result}(b). We demonstrate using the open Dicke model (DM) \cite{emary_chaos_2003, dimer_proposed_2007} and its one-dimensional lattice extension, the open Dicke lattice model (DLM) \cite{zou_implementation_2014} that the range of $\tau_{q}$ where we observe this ``delayed" KZM is a generic feature of open systems with finite dissipation strength, $\kappa$. We also show that the delayed KZM is more prominent near criticality and that its signatures become more significant for large threshold values for an order parameter. Thus, our paper highlights subtleties of the KZM in open systems in finite quench scenarios relevant to experiments.

The paper is structured as follows. In Sec.~\ref{sec: theory}, we introduce a minimal system that can exhibit the delayed KZM when quenched at intermediate values of $\tau_{q}$. By deriving an effective potential for the minimal system, we show that the phenomenon is due to a relaxation mechanism induced by the dissipation of the system. Then, using the open DM and open DLM as a test bed, we demonstrate in Sec.~\ref{sec: spin_cavity_kzm} that the delayed KZM is a generic feature of open systems under a finite quench and that the phenomenon becomes more prominent near criticality. In Sec.~\ref{sec: transition_time}, we explore how the deviations brought by the delayed KZM can be further exacerbated with large thresholds for order parameters and propose an alternative method for measuring the transition time beyond the threshold-based criterion. We provide a summary and possible extensions of our work in Sec.~\ref{sec: conclusion}.

\section{Delayed KZM: Theory}\label{sec: theory}

\begin{figure}
	\centering
	\includegraphics[scale=0.5]{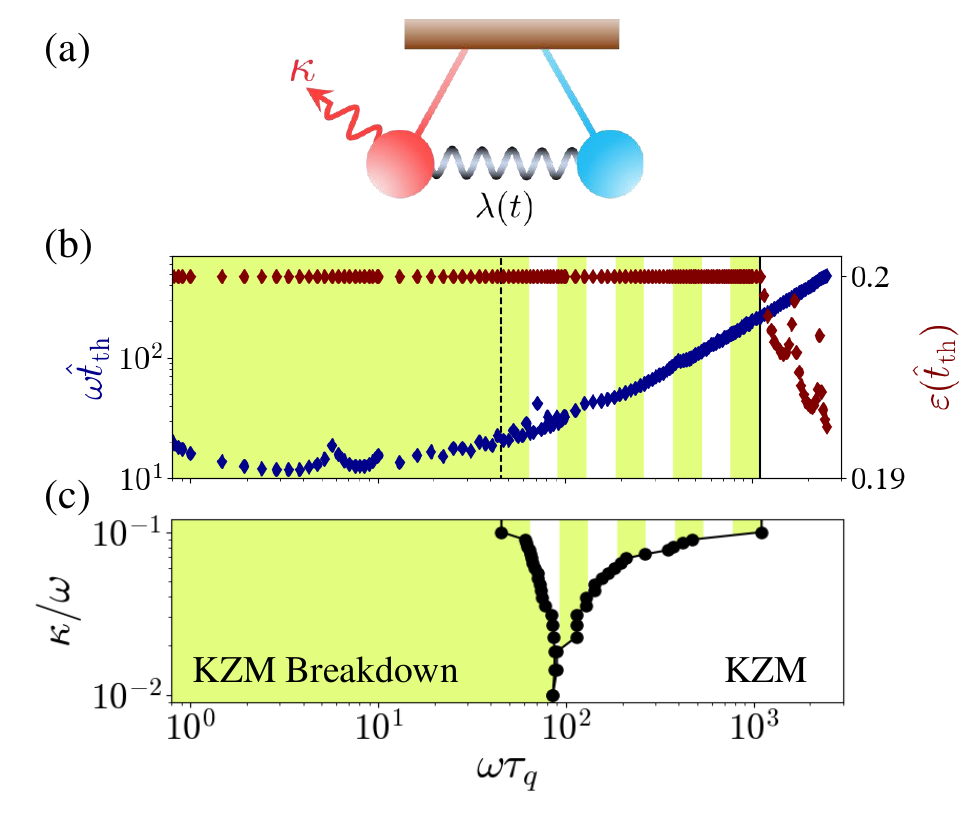}
	\caption{(a) Sketch of the open COS. (b) Scaling of $\hat{t}_{\mathrm{th}}$ and $\varepsilon(\hat{t}_{\mathrm{th}})$ as a function of $\tau_{q}$ for $\kappa = 0.1 \omega$. The vertical dashed line corresponds to $\tau_{q, c}$, while the solid line corresponds to $\tau_{q, c}^{*}$. (c) Boundary of the delayed KZM as a function of $\kappa$. The remaining parameters are set to $\varepsilon_{f} = 0.2$ and $|a|_{\mathrm{th}}^{2} = 2$.}
	\label{fig: cos_delayed_kzm}
\end{figure}

Consider a generic open system with a continuous phase transition that is described by the Lindblad master equation \cite{breuer_open_systems_2002},
\begin{equation}
\label{eq:master_equation}
\partial_{t} \hat{\rho} = -i \left[ \hat{H} \left( \varepsilon(t) \right) / \hbar, \hat{\rho} \right] +  \mathcal{D} \hat{\rho},
\end{equation}
where $\mathcal{D} \hat{\rho} = \sum_{\ell} \kappa_{\ell} \left( 2 \hat{L}_{\ell} \hat{\rho} \hat{L}_{\ell}^{\dagger} - \left\{ \hat{L}^{\dagger}_{\ell}\hat{L}_{\ell}, \hat{\rho} \right\} \right)$ is the dissipator and $\hat{H}(\varepsilon(t))$ is the time-dependent Hamiltonian of the system. The system undergoes a phase transition from a normal phase (NP), in which the global symmetry of the system is preserved, to a symmetry-broken phase via a finite quench protocol,
\begin{equation}
\label{eq:finite_ramp_protocol}
    \varepsilon(t) = 
    \begin{cases}
        t / \tau_{q} \; \; \;  & t_{i} \leq t \leq \varepsilon_{f}\tau_{q} \\
        \varepsilon_{f} \; \; \;  & \varepsilon_{f}\tau_{q} < t \leq t_{f}
    \end{cases}
\end{equation}
where $t_{i} = -\tau_{q}$ and $t_{f}$ are the initial and final time of the quench. In the following, we demonstrate that if the system can be approximated as or mapped onto an effective coupled oscillator system (COS), with at least one dissipative channel, as sketched in Fig.~\ref{fig: cos_delayed_kzm}(a), then we should observe a finite range of $\tau_{q}$ where the deviation between $\hat{t}$ and $\hat{t}_{\mathrm{th}}$ becomes significant enough that we get a contradictory behavior between the scaling of the transition time and crossover point. To observe the dynamics of the systems considered in this paper, we will use a mean-field approach and assume that for any operators, $\hat{A}$ and $\hat{B}$, $\left< \hat{A} \hat{B} \right> \approx \left< \hat{A} \right> \left< \hat{B} \right>$. This allows us to treat any operators as complex numbers and use the notation $A \equiv \left< \hat{A} \right>$. We numerically integrate the systems' mean-field equations in Appendix \ref{sec: mean_field_eom} using a standard fourth-order Runge Kutta algorithm with a time step of $\omega \triangle t = 0.01$, where $\omega$ is a frequency associated to the dissipative channel, as we will show later.

The Hamiltonian of the COS is 
\begin{equation}
\label{eq: cos_hamiltonian}
\frac{\hat{H}^{\mathrm{COS}}}{\hbar} = \omega \hat{a}^{\dagger}\hat{a} + \omega_{0}\hat{b}^{\dagger}\hat{b} + \lambda(t) \left( \hat{a}^{\dagger} + \hat{a} \right) \left( \hat{b} + \hat{b}^{\dagger} \right) ,
\end{equation}
where $\omega$ and $\omega_{0}$ are the transition frequencies associated with the bosonic modes $\hat{a}$ and $\hat{b}$, respectively, and $\lambda(t)$ is the coupling strength between the two modes. The $\hat{a}$ mode is subject to dissipation, which is captured in the master equation by the dissipator $\mathcal{D}\hat{\rho} = \kappa \left(2 \hat{a}\hat{\rho} \hat{a}^{\dagger} - \left\{ \hat{a}^{\dagger}\hat{a}, \hat{\rho} \right\} \right)$. The COS has an extensive application in multiple settings, such as---but not limited---to cavity-magnon systems \cite{rameshti_cavity_2022, kim_observation_2024}, atom-cavity systems \cite{emary_chaos_2003, dimer_proposed_2007, skulte_realizing_2024, jara_theory_2024}, and spin systems \cite{calvanese_strinati_theory_2019, bello_persistent_2019}.

The open COS has two phases: the NP which corresponds to a steady state with $a = b = 0$, and an unbounded state where both modes exponentially diverge as $t \rightarrow \infty$ \cite{emary_chaos_2003}. When nonlinearity is present, the unbounded state can be associated with a symmetry-broken phase, in which the modes choose a new steady state depending on their initial values. These two states are separated by the critical point \cite{emary_chaos_2003}:
\begin{equation}
\label{eq: cos_critical_point}
\lambda_{c} = \frac{1}{2} \sqrt{\frac{\omega_{0}}{{\omega}} \left( \kappa^{2} + \omega^{2} \right) } .
\end{equation}

To show that the COS is a minimal model that can exhibit the KZM and its subsequent breakdown at small $\tau_{q}$, we consider its dynamics as it transitions from the NP to the unbounded state. We do this by initializing the system near the steady state of NP, $a_{0}=-b_{0}=0.01$. We then apply the quench protocol in Eq.~\eqref{eq:finite_ramp_protocol} onto the COS and track the dynamics of the occupation number of the $\hat{a}$ mode, $|a|^{2}$. We finally determine $\hat{t}_{\mathrm{th}}$ by identifying the time it takes for $|a|^{2}$ to reach the threshold value, $|a|^{2}_{\mathrm{th}}$, after the ramp passes $\varepsilon(t = 0)=0$. The crossover point at the transition time is then inferred back from $\hat{t}_{\mathrm{th}}$ using Eq.~\eqref{eq:finite_ramp_protocol}. We present in Fig.~\ref{fig: cos_delayed_kzm}(b) the scaling of $\hat{t}_{\mathrm{th}}$ and $\varepsilon(\hat{t}_{\mathrm{th}})$ as a function of $\tau_{q}$. We can observe that for large $\tau_{q}$, or slow quench, all relevant quantities follow the power-law scaling predicted by the KZM. As we decrease $\tau_{q}$, $\varepsilon(\hat{t}_{\mathrm{th}})$ begins to saturate at a larger critical quench time, $\tau_{q, c}^{*}$, than $\hat{t}_{\mathrm{th}}$, as indicated by the solid line in Fig.~\ref{fig: cos_delayed_kzm}(b). Finally, as $\tau_{q}\rightarrow 0$, $\hat{t}_{\mathrm{th}}$ approaches a constant value after passing another critical quench time, $\tau_{q, c}$, denoted in Fig.~\ref{fig: cos_delayed_kzm}(b) as a dashed line. Note that the fluctuations in the scaling of $\hat{t}_{\mathrm{th}}$ and $\varepsilon(\hat{t}_{\mathrm{th}})$ can be attributed to the mean-field approach, which neglects any quantum fluctuation in the system's dynamics. The scaling behavior of $\hat{t}_{\mathrm{th}}$ and $\varepsilon(\hat{t}_{\mathrm{th}})$ implies that within the range $\tau_{q, c} < \tau_{q} \leq \tau_{q, c}^{*}$, there exists an intermediate regime between the true breakdown and the validity of the KZM, wherein the KZM remains valid even though the system appears to relax well after the quench has terminated. As shown in Fig.~\ref{fig: cos_delayed_kzm}(c), this intermediate regime vanishes as $\kappa \rightarrow 0$, highlighting that this is a dissipation-induced effect.

\begin{figure}
	\centering
	\includegraphics[scale=0.40]{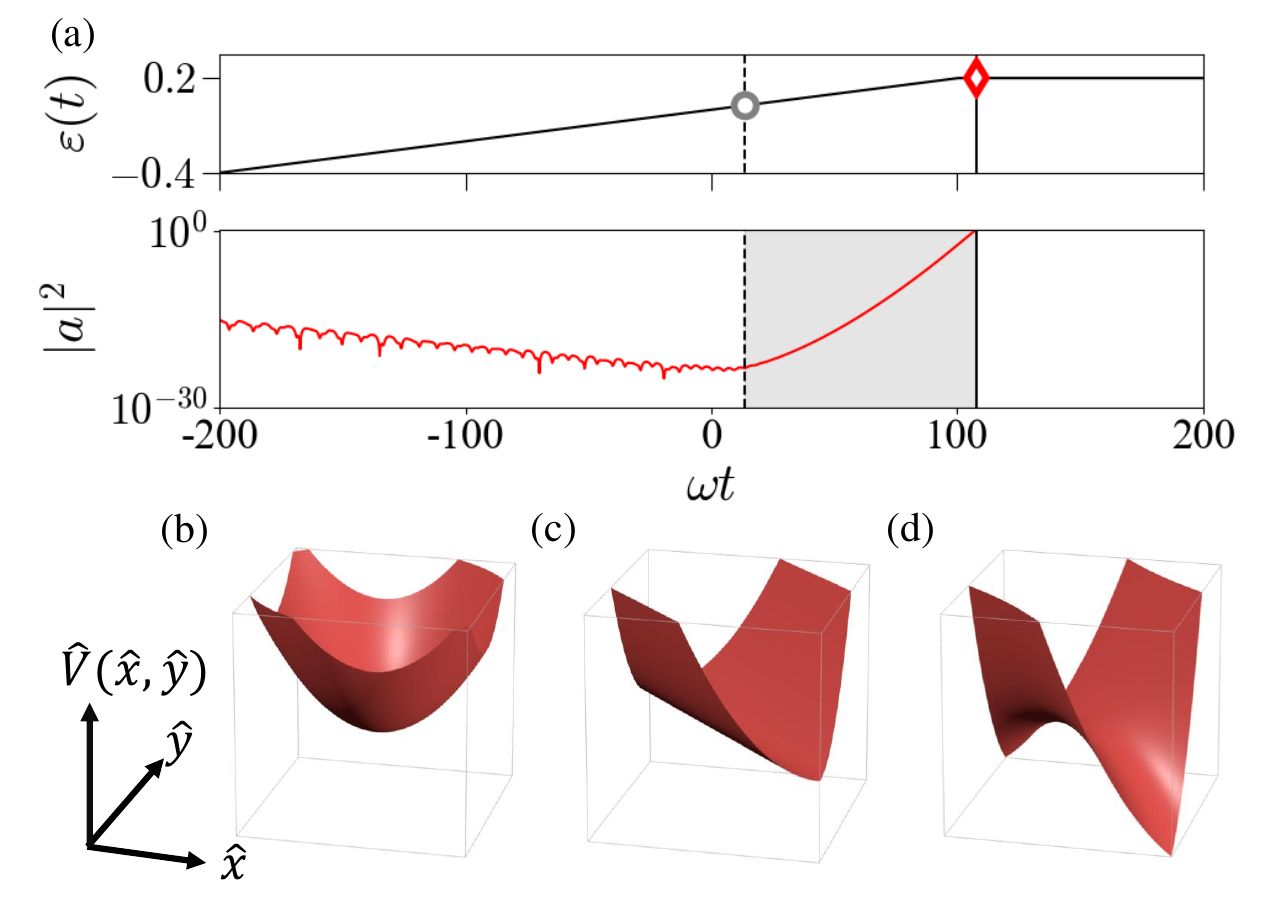}
	\caption{(a) (top panel) Ramp protocol for $\omega \tau_{q} = 500$ and $\varepsilon_{f} = 0.2$. The circle point indicates $\varepsilon(\hat{t})$ while the diamond point marks $\varepsilon(\hat{t}_{\mathrm{th}})$. Bottom panel: Exemplary dynamics of $|a|^{2}$ in the $\tau_{q}$-regime of the delayed KZM for $\kappa = 0.1 \omega$. The vertical dashed lines correspond to $\omega \hat{t}$, while the solid lines represent $\omega \hat{t}_{\mathrm{th}}$ for a threshold of $|a|^{2}_{\mathrm{th}}= 2$. (b), (c) potential surface of the open COS for (b) $\varepsilon(t) < 0$, (c) $\varepsilon(t) = 0$, and (d) $\varepsilon(t) > 0$.  }
	\label{fig: cos_potential_surface}
\end{figure}

We can understand this apparent contradiction between the scaling of the $\hat{t}_{\mathrm{th}}$ and $\varepsilon(\hat{t}_{\mathrm{th}})$ by looking at the dynamics of $|a|^{2}$ as the ramp crosses over $\varepsilon = 0$. In Fig.~\ref{fig: cos_potential_surface}(a), we present an exemplary dynamics of $|a|^{2}$ in the logarithmic scale for the regime $\tau_{q, c} < \tau_{q} \leq \tau_{q, c}^{*}$. Notice that before the system enters the unbounded state, $|a|^{2}$ first exponentially decays towards its steady state, indicating that the system does not freeze in the impulse regime. This dynamics is reminiscent of systems relaxing towards the global minimum of their energy surface due to dissipation, as sketched in Fig.~\ref{fig: sketch_result}(b). We can further establish this connection by obtaining the potential surface of the COS, which we can do by substituting the pseudoposition and momentum operators for the $\hat{a}$ mode,
\begin{equation}
\label{eq: a_position_momentum}
\hat{x} = \frac{1}{\sqrt{2\omega}} \left(\hat{a}^{\dagger} + \hat{a} \right), \quad \hat{p}_{x} = i\sqrt{\frac{\omega}{2}} \left(\hat{a}^{\dagger} - \hat{a}  \right),
\end{equation}
and the $\hat{b}$ mode,
\begin{equation}
\label{eq: b_position_momentum}
\hat{y} = \frac{1}{\sqrt{2\omega_{0}}} \left(\hat{b}^{\dagger} + \hat{b} \right), \quad \hat{p}_{y} = \sqrt{\frac{\omega_{0}}{2}} \left( \hat{b}^{\dagger} - \hat{b} \right),
\end{equation}
back to Eq.~\eqref{eq: cos_hamiltonian}. Note that for the remainder of this section, we set $\hbar = 1$ for brevity. With this substitution, the COS Hamiltonian becomes
\begin{equation}
\label{eq: cos_cartesian}
\hat{H}^{\mathrm{COS}} = \frac{\hat{p}_{x}^{2}}{2} + \frac{\hat{p}_{y}^{2}}{2} + \hat{V}\left(\hat{x}, \hat{y}\right),
\end{equation}
where
\begin{equation}
\label{eq: cos_potential}
\hat{V}\left(\hat{x}, \hat{y}\right) = \frac{1}{2} \omega^{2} \hat{x}^{2} + \frac{1}{2} \omega_{0}^{2} \hat{y}^{2} + 2 \sqrt{\omega \omega_{0}} \lambda \hat{x} \hat{y}
\end{equation}
is the effective potential of the COS in the closed system limit, $\kappa = 0$. In this limit, the potential surface has a global minimum at $\hat{x} = \hat{y} = 0$ when $\lambda < \lambda_{c} = \sqrt{\omega \omega_{0}} / 2$, as shown in Fig.~\ref{fig: cos_potential_surface}(b). It then loses its global minimum when $\lambda = \lambda_{c}$ as sketched in Fig.~\ref{fig: cos_potential_surface}(c). Finally, the global minimum becomes a saddle point when $\lambda > \lambda_{c}$, as shown in Fig.~\ref{fig: cos_potential_surface}(d). Note that in the presence of dissipation, the COS effective potential only becomes modified such that the critical point becomes Eq.~\eqref{eq: cos_critical_point}, while the structure of the potential surface remains the same due to Eq.~\eqref{eq: cos_cartesian} being quadratic.

With the above picture, we can now interpret the relaxation mechanism observed in Fig.~\ref{fig: cos_potential_surface}(a) as follows. Suppose that we initialize our system such that $\lambda < \lambda_{c}$ and the initial states of $\hat{a}$ and $\hat{b}$ modes are close to the global minimum of $\hat{V}$. In the mean-field level, if $\kappa = 0$, we can expect that the system will oscillate around the global minimum of $\hat{V}$ as we increase $\lambda$ using the finite ramp protocol defined in Eq.~\eqref{eq:finite_ramp_protocol}, together with the modification of the COS potential surface. As we cross $\lambda_{c}$, the global minimum of $\hat{V}$ becomes a saddle point. As such, any deviation of the initial state from the origin would eventually push the system to either the positive $\hat{x}$ and $-\hat{y}$ direction or vice versa, signaling the spontaneous symmetry breaking of the system.

In the presence of dissipation, however, the system can still relax to the global minimum before the quench reaches $\lambda_{c}$ for sufficiently large quench timescales $\tau_{q} > \tau_{q, c}$. As a result, the slow deformation of the effective potential allows for the system to remain near $a = b = 0$ even after passing the critical point where the potential loses its global minimum. The nudge from the system's initial state eventually pushes the system towards a new minimum as the quench progresses, signaling the phase transition. This approach, however, only becomes detectable when $|a|^{2}$ reaches $|a|_{\mathrm{th}}^{2}$, which occurs only after the linear ramp has terminated. Thus, we observe the saturation of the crossover point at $\varepsilon_{f}$ even though $\hat{t}_{\mathrm{th}}$ follows the predicted scaling of the KZM, which hints that the system entered the adiabatic regime within the duration of the ramp. Note that the relaxation mechanism is not present in the closed limit, as hinted by the regime vanishing in Fig.~\ref{fig: cos_delayed_kzm}(c) as $\kappa \rightarrow 0$. The delay between $\hat{t}$ and $\hat{t}_{\mathrm{th}}$ at finite $\tau_{q}$ motivates us to call this phenomenon delayed KZM.

In the next section, we will show that the delayed KZM is a generic feature of open systems that can be mapped onto an effective COS. Moreover, we will demonstrate that not only is the delayed KZM induced purely by dissipation but it also becomes more prominent when the system is quenched near criticality, $\varepsilon_{f} \approx 0$.

\section{Delayed KZM in open systems}\label{sec: spin_cavity_kzm}

\subsection{Signatures of the delayed KZM} \label{subsec: delayed_kzm_signatures}

\begin{figure*}[!htbp]
	\centering
	\includegraphics[scale=0.525]{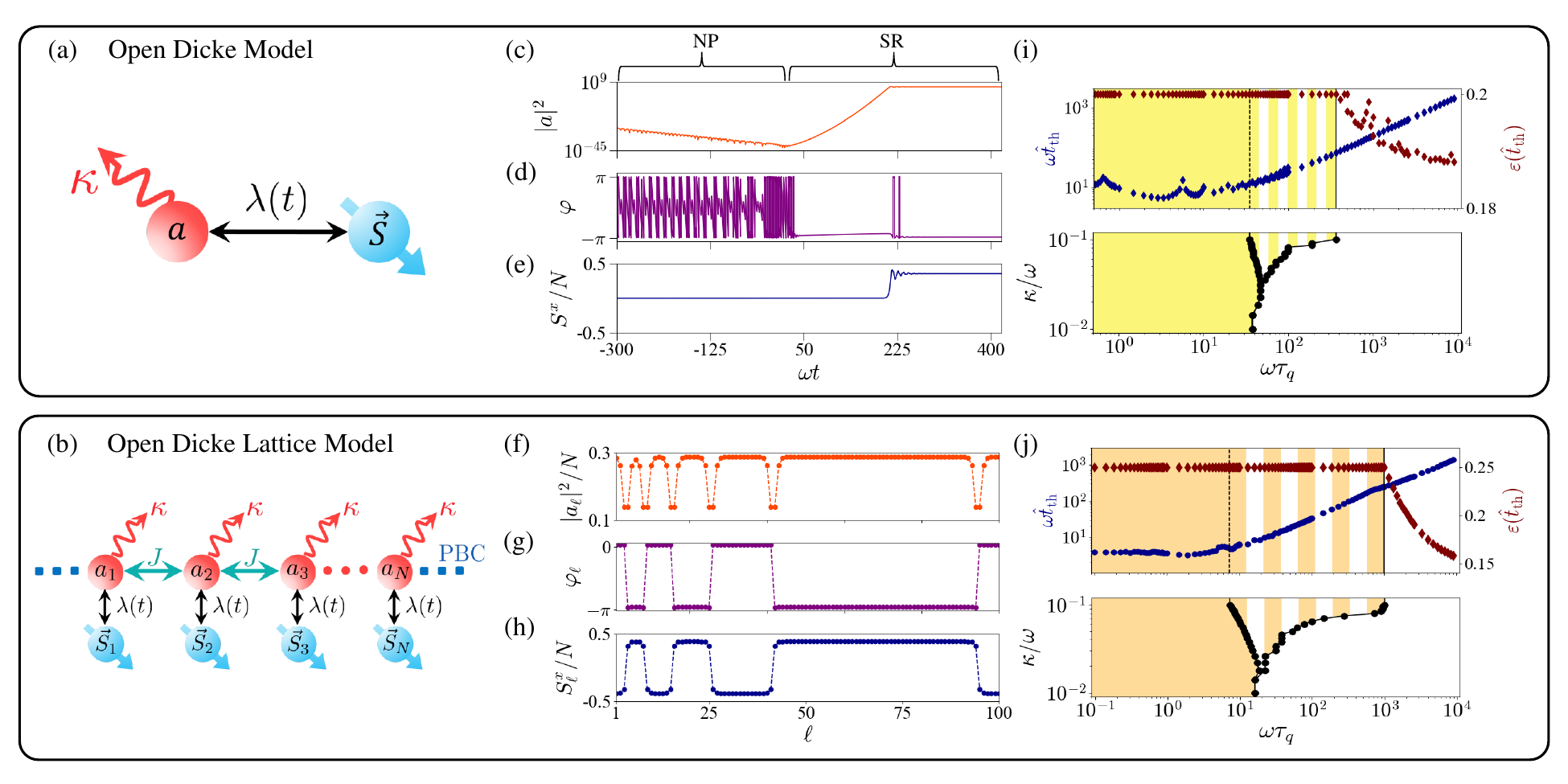}
	\caption{(a), (b) Sketch of the (a) open Dicke model and the (b) open Dicke lattice model with periodic boundary condition. (c)-(e) Exemplary dynamics of the (c) occupation number, (d) phase of the $\hat{a}$ mode, and (e) $S^{x}$ of the quenched open Dicke model for $\omega \tau_{q} = 1000$ and $\varepsilon_{f} = 0.2$. (f)-(h) Exemplary steady-state spatial distribution of the (f) occupation number, (g) phase of the $\hat{a}_{\ell}$ mode, and (h) $S^{x}_{\ell}$ of the open Dicke lattice model in the SR phase. (i)-(j) Signatures of the delayed KZM for the (i) open DM and (j) open DLM. Top panel: Scaling of $\hat{t}_{\mathrm{th}}$ as a function of $\tau_{q}$ for $\kappa = 0.1\omega$. The vertical dashed lines represent $\tau_{q, c}$, while the solid lines corresponds to $\tau_{q, c}^{*}$. Bottom panel: The boundary of the delayed KZM  regime as a function of $\kappa$. The remaining parameters are set to $\varepsilon_{f} = 0.2$ and $|a|^{2}_{\mathrm{th}} = 2$ for the open DM, and $J = 0.1\omega$, $\varepsilon_{f} = 0.25$, and $|a|^{2}_{\mathrm{th}} / M = 2$, with $M = 500$ sites, for the open DLM. }
	\label{fig: open_dm_and_dlm}
\end{figure*}

We now test whether the delayed KZM is a generic feature of open systems by considering two fully connected systems: the open DM, schematically represented in Fig.~\ref{fig: open_dm_and_dlm}(a), and its one-dimensional lattice version, the open DLM, as shown in Fig.~\ref{fig: open_dm_and_dlm}(b). Both systems are described by the master equation in Eq.~\eqref{eq:master_equation}, with the Hamiltonian of the open DM being \cite{emary_chaos_2003, dimer_proposed_2007} \begin{equation}
\label{eq:open_dm_hamiltonian}
\frac{\hat{H}^{\mathrm{DM}}}{\hbar} = \omega \hat{a}^{\dagger}\hat{a} + \omega_{0}\hat{S}^{z} + \frac{2\lambda(t)}{\sqrt{N}}\left( \hat{a} + \hat{a}^{\dagger} \right) \hat{S}^{x},
\end{equation} 
while the Hamiltonian of its $M$-site lattice version with periodic boundary conditions takes the form \cite{zou_implementation_2014}
\begin{equation}
\label{eq: open_dlm_hamiltonian}
\frac{\hat{H}^{\mathrm{DLM}}}{\hbar} = \frac{1}{\hbar}\sum_{\ell}^{M} \hat{H}^{\mathrm{DM}}_{\ell} - J \sum_{\left<i, j \right>}^{M}  \left( \hat{a}^{\dagger}_{i}\hat{a}_{j} + \hat{a}^{\dagger}_{j}\hat{a}_{i}  \right) .
\end{equation}
The open DM has the same dissipator as the open COS, while the dissipator of the open DM is $\mathcal{D} \hat{\rho} = \kappa \sum_{\ell}^{M} \left( 2\hat{a}_{\ell}\hat{\rho}\hat{a}_{\ell}^{\dagger} - \left\{ \hat{a}_{\ell}^{\dagger} \hat{a}_{\ell}, \hat{\rho} \right\} \right) $ \cite{zou_implementation_2014}. The open DM describes the dynamics of $N$ two-level systems, represented by the collective spin operators $\hat{S}^{x, y, z}$, coupled to a dissipative bosonic mode, $\hat{a}$, which in cavity-QED experiments corresponds to a photonic mode \cite{klinder_dynamical_2015, baumann_dicke_2010, dimer_proposed_2007, skulte_realizing_2024}. In both systems, $\omega$ and $\omega_{0}$ are the bosonic and spin transition frequencies, respectively, and $\lambda$ is the spin-boson coupling, while $J$ represents the nearest-neighbor interaction in the open DLM.

In equilibrium, the open DM has two phases: the NP and the superradiant (SR) phase \cite{emary_chaos_2003, dimer_proposed_2007}. The NP is characterized by a fully polarised collective spin at the $-z$ direction, i.e. $S^{z} = -N / 2$, and a zero total occupation number, $|a|^{2}$. Meanwhile, the SR phase is associated with the $\mathbb{Z}_{2}$ symmetry breaking of the system, leading to a nonzero $S^{x}$ and $|a|^{2}$, with $S^{x}$ ($a$) picking a random sign (phase) from the two degenerate steady states of the system \cite{dimer_proposed_2007}. The two phases are separated by the same critical point as the open COS \cite{dimer_proposed_2007}. Under a finite quench, however, the open DM exhibits nontrivial dynamics as it transitions from the NP to the SR phase. We present in Figs.~\ref{fig: open_dm_and_dlm}(c)-\ref{fig: open_dm_and_dlm}(e) an exemplary dynamics of the total occupation number, $|a|^{2}$, and the phase of the bosonic mode, $\varphi$, and $S^{x}$ of the open DM for $\omega \tau_{q} = 1000$. We initialized the system near the steady state of the NP, where the initial values of the bosonic mode are $a=0.01$, while the collective spin operators are 
\begin{equation}
\label{eq: spin_initial_state}
\quad S^{x}(t_{i}) = \frac{N}{2}\delta, \quad S^{y}(t_{i}) =0, \quad S^{z}(t_{i})= -\frac{N}{2}\sqrt{1 - \delta^{2}},
\end{equation}
where $\delta$ is a perturbation set to $\delta = 0.01$. We can observe that when the system is in the NP, the occupation number approaches the NP steady state, $a=0$, which is consistent with our predicted behavior from the potential surface interpretation of phase transition in an open system, which we describe in Sec.~\ref{sec: theory}. In addition, the phase of the bosonic mode oscillates from $-\pi$ to $\pi$, while the $S^{x}$ remains close to $S^{x}=0$. As $\varepsilon(t)>0$ at $t>0$, the system enters the SR phase, which results in $\varphi$ spontaneously admitting a finite value as the $|a|^{2}$ starts to exponentially grow until $t = \varepsilon_{f}\tau_{q}$, where the ramp terminates. At that point, the $|a|^{2}$ finally saturates at the steady state of the SR phase. Meanwhile, the transition of $S^{x}$ from its behavior in the NP to the SR phase only becomes prominent at a later time. We will further expand on the implication of the behavior of these order parameters later in Sec.~\ref{sec: transition_time}.

As for the open DLM, for small values of $J$, the interaction between the open DMs modifies the $\lambda_{c}$ into a critical line \cite{zou_implementation_2014}: 
\begin{equation}
\label{eq: open_dm_critical_line}
\lambda_{c} = \frac{1}{2} \sqrt{\omega_{0} (\omega - 2J) \left( 1 + \frac{\kappa^{2}}{(\omega - 2J)^{2}} \right)}.
\end{equation} 
Moreover, suppose that we drive the open DLM from the NP to the SR phase using a finite quench after initializing it near the steady state of NP. Specifically, we initialize the collective spins at $S^{x, y, z}_{\ell} = S^{x, y, z}(t = t_{i}) $, while the bosonic modes are initialized at the vacuum state, which can be represented as a complex Gaussian variable $a_{\ell} = \frac{1}{2} \left( \eta_{\ell}^{\mathrm{R}} + \eta_{\ell}^{\mathrm{I}} \right)$, where $\eta_{\ell}^{R, I}$ are random numbers sampled from a Gaussian distribution satisfying $\left< \eta_{\ell}^{i} \right> = 0$ and $\left< \eta_{\ell}^{i} \eta_{m}^{j}  \right> = \delta_{i, j}\delta_{\ell, m}$ for $i, j = R, I$ \cite{olsen_numerical_2009}. Then, as the system enters the SR phase, each site can independently pick between the two degenerate steady states available, allowing for the formation of domains and point defects, the number of which depends on the correlation length of the system. We present in Figs.~\ref{fig: open_dm_and_dlm}(f)-\ref{fig: open_dm_and_dlm}(h) the exemplary spatiotemporal dynamics of $|a_{\ell}|^{2}$, $\varphi_{\ell}$, and $S^{x}_{\ell}$ of the open DLM after doing a finite quench towards the SR phase. We can observe that the point defects can manifest either as dips in the occupation number, phase slips in the spatial profile of $\varphi_{\ell}$, or domain walls in $S^{x}_{\ell}$. Note that the defect number $N_{d}$ follows the predicted KZM power-law scaling with $\tau_{q}$, which we demonstrate in Appendix \ref{sec: kzm_exponents}. Since the notion of topological defects is well-defined in the open DLM, it serves as a good test bed for the delayed KZM for systems with short-range interaction. This is in addition to the open DM, which has been experimentally shown to exhibit signatures of the KZM \cite{klinder_dynamical_2015} despite the open question of its nonequilibrium universality class \cite{acevedo_new_2014, caneva_adiabatic_2008}.

We now present in Figs.~\ref{fig: open_dm_and_dlm}(i) and \ref{fig: open_dm_and_dlm}(j) the scaling of $\hat{t}_{\mathrm{th}}$ and $\varepsilon(\hat{t}_{\mathrm{th}})$ as a function of $\tau_{q}$ for the open DM and open DLM, respectively. Similar to the COS, the $\hat{t}_{\mathrm{th}}$ for both systems is inferred from the total occupation number, which for the open DLM is explicitly defined as $|a|^{2} = \sum_{\ell} |a_{\ell}|^{2}$. Notice that both systems exhibit the signatures of the delayed KZM, where $\hat{t}_{\mathrm{th}}$ continues with its KZM power-law scaling as $\varepsilon(\hat{t}_{\mathrm{th}})$ saturates for intermediate values of $\tau_{q}$. They also exhibit the closing of the boundary of the delayed KZM as we decrease $\kappa$. We can understand the emergence of the delayed KZM in these two systems by noting that the open DM can be mapped exactly into the COS in the thermodynamic limit, $N \rightarrow \infty$. We can do this by applying the approximate Holstein-Primakoff representation (HPR) \cite{emary_chaos_2003, dimer_proposed_2007},
\begin{equation}
\label{eq: hpr_approx}
\hat{S}^{z} = \frac{N}{2}, \quad \hat{S}^{-}= \sqrt{N} \left( \sqrt{1 - \frac{\hat{b}^{\dagger}\hat{b} }{N}} \right) \hat{b}\approx \sqrt{N}\hat{b},
\end{equation} 
on Eq.~\eqref{eq:open_dm_hamiltonian} to reduce it onto the COS Hamiltonian in Eq.~\eqref{eq: cos_hamiltonian} up to a constant term.

Meanwhile, we can transform the open DLM into a set of COS in the thermodynamic limit by first substituting the approximate HPR of the collective spins to Eq.~\eqref{eq: open_dlm_hamiltonian}, noting that $\hat{S}^{z, \pm} \rightarrow \hat{S}^{z, \pm}_{\ell}$ and $\hat{b} \rightarrow \hat{b}_{\ell}$ 
\cite{zou_implementation_2014}. This leads to a Hamiltonian of the form,
\begin{equation}
\label{eq: coupled_cos}
\frac{\hat{H}^{\mathrm{DLM}}}{\hbar} \approx \frac{1}{\hbar} \sum_{\ell}^{M} \hat{H}_{\ell}^{\mathrm{COS}} - J \sum_{\left<i, j \right>} \left( \hat{a}_{i}^{\dagger}\hat{a}_{j} + \hat{a}_{j}^{\dagger} \hat{a}_{i} \right).
\end{equation}
We then perform a discrete Fourier transform,
\begin{equation}
\hat{a}_{k} = \frac{1}{\sqrt{M}} \sum_{\ell}e^{ik\ell} \hat{a}_{\ell}, \quad \hat{b}_{k} = \frac{1}{\sqrt{M}} \sum_{\ell} e^{ik\ell} \hat{b}_{\ell},
\end{equation} 
on Eq.~\eqref{eq: coupled_cos} to obtain an effective Hamiltonian,
\begin{equation}
\label{eq: uncoupled_dlm}
\frac{\hat{H}^{\mathrm{DLM}}}{\hbar} \approx \frac{1}{\hbar} \sum_{k} \hat{H}^{\mathrm{OM}}_{k}, 
\end{equation}
where
\begin{equation}
\frac{\hat{H}^{\mathrm{OM}}_{k}}{\hbar} = \omega_{k} \hat{a}_{k}^{\dagger}\hat{a}_{k} + \omega_{0}\hat{b}_{k}^{\dagger}\hat{b}_{k} + \lambda \left( \hat{a}_{k}^{\dagger}\hat{b}_{k} + \hat{a}_{-k}\hat{b}_{k} + \mathrm{H.c.} \right)
\end{equation}
is the Hamiltonian of each uncoupled oscillator at the momentum mode $k$ and $\omega_{k} = \omega - 2J\cos(k)$. In this form, we can easily observe that the open DLM has a similar structure to the COS, with the similarity being more apparent at the zero-momentum mode:
\begin{equation}
\frac{\hat{H}^{\mathrm{OM}}_{0}}{\hbar}= \left( \omega - 2J \right)\hat{a}_{0}^{\dagger}\hat{a}_{0} + \omega_{0}\hat{b}_{0}^{\dagger}\hat{b}_{0} + \lambda \left(\hat{a}_{0}^{\dagger} + \hat{a}_{0} \right)\left(\hat{b}_{0}^{\dagger} + \hat{b}_{0} \right).
\end{equation}
These results show that the signatures of the delayed KZM can appear not only in the open DM but also in the open DLM, where both short-range interactions between the sites and multiple degenerate steady states are present in the system. As such, we confirm that the delayed KZM is a generic feature of open systems under a finite quench that can be mapped onto a COS, regardless of the interaction present in the system.

Since we have shown the generality of the delayed KZM on open systems, we now explore in greater detail the dissipative and near-critical nature of the delayed KZM in Sec.~\ref{subsec: dissipative_delayed_kzm}.

\subsection{Dissipative and critical nature of the delayed KZM}\label{subsec: dissipative_delayed_kzm}

In Sec.~\ref{sec: theory}, we have claimed that the dissipation is responsible for the relaxation mechanism that leads to the emergence of the delayed KZM. This is also corroborated by the disappearance of the delayed KZM regime in the closed limit, implying that the phenomenon appears only at finite dissipation strength, $\kappa$. We now explicitly demonstrate that this claim is true for any generic open systems by calculating the decay rate of the total occupation number, $\gamma_{d}$, as the system approaches the critical point. We will then identify how $\gamma_{d}$ scales with $\varepsilon_{f}$ and $\kappa$. For the rest of this section, we will only consider the open DLM, although our results here should apply as well for both the COS and the open DM.

\begin{figure}
	\centering
	\includegraphics[scale=0.28]{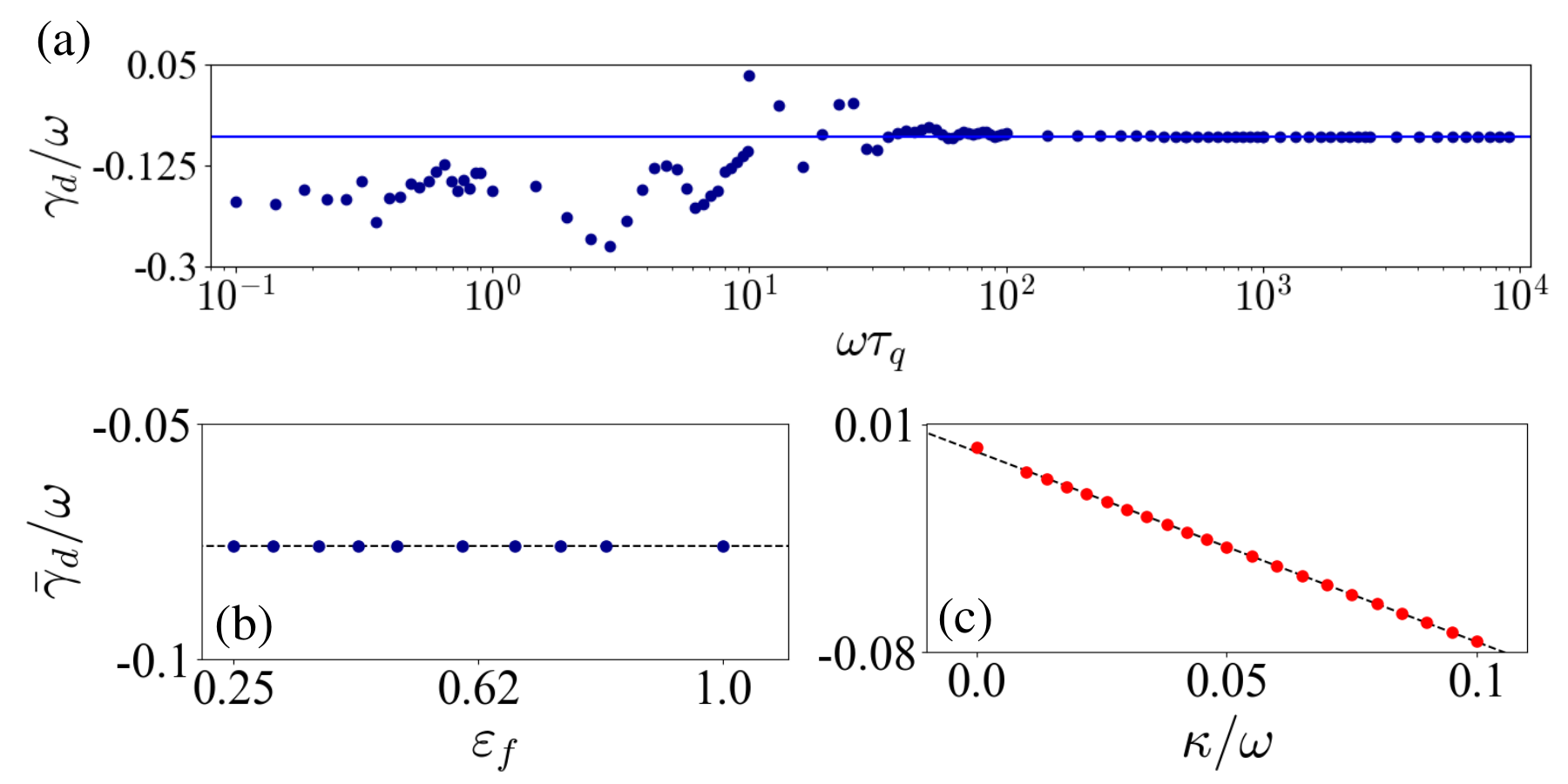}
	\caption{(a) Dependence of the occupation number decay rate of the open DLM on $\tau_{q}$ for $\varepsilon_{f} = 0.25$ and $\kappa = J = 0.1\omega$. The solid line corresponds to the average value of $\gamma_{d}$ from $\omega \tau_{q}=10^{2}$ to $\omega \tau_{q} = 10^{4}$. (b), (c) $\bar{\gamma}_{d}$ as a function of (b) $\varepsilon_{f}$ for $\kappa = 0.1\omega$ and (c) $\kappa$ for $\varepsilon_{f} = 0.25$. The dashed lines correspond to the best-fit lines. }
	\label{fig:dissipation_delayed_kzm}
\end{figure}

To determine the $\gamma_{d}$ of the open DLM for a given $\kappa$ and $\varepsilon_{f}$, we calculate the slope of the best-fit line of the logarithm of $|a|^{2}$ within the time interval $\left[-0.75 \tau_{q}, 0\right]$. The chosen time window is arbitrary, but it ensures that the $\gamma_{d}$ is inferred within the duration that the system is in the impulse regime. We show in Fig.~\ref{fig:dissipation_delayed_kzm}(a) the dependence of $\gamma_{d}$ with the quench time. We can observe that $\gamma_{d}$ is constant for large values of $\tau_{q}$. As we decrease $\tau_{q}$, however, $\gamma_{d}$ begins to fluctuate and eventually decreases to a much lower value. We attribute the deviation of $\gamma_{d}$ from its constant value on the errors incurred in the best-fit line of $\ln|a|^{2}$ for small values of $\tau_{q}$. In particular, since we only considered a simulation time step of $\omega \triangle t = 0.01$, the small time window for these values of $\tau_{q}$ leads to smaller sets of data points for $|a|^{2}$, resulting to an overall poorer fit. Due to this consideration, we only considered the data points from $\omega \tau_{q} = 10^{2}$ to $\omega \tau_{q} = 10^{4}$ in calculating the average value of the decay rate with $\tau_{q}$, $\bar{\gamma}_{d}$.

We now present in Figs.~\ref{fig:dissipation_delayed_kzm}(b) and \ref{fig:dissipation_delayed_kzm}(c) the behavior of $\bar{\gamma}_{d}$ as a function of $\varepsilon_{f}$ and $\kappa$, respectively. We can observe that $\bar{\gamma}_{d}$ remains constant for all values of $\varepsilon_{f}$, implying that the average decay rate of $|a|^{2}$ is independent of the quench protocol used in the system. Meanwhile, $\bar{\gamma}_{d}$ has an inverse relationship with $\kappa$, demonstrating that the relaxation mechanism responsible for the delayed KZM is indeed a direct result of dissipation allowing the initial occupation number in the dissipative bosonic mode to leak out of the system as it remains in the impulse regime. For completeness, we check the linear dependence of $\bar{\gamma}_{d}$ with $\kappa$ by fitting a line on it and calculating the square of its Pearson correlation coefficient, $R^{2}$. By doing this, we obtain $R^{2} = 0.9996$, which indicates a great fit between the best-fit line and the data points.

\begin{figure}
	\centering
	\includegraphics[scale=0.45]{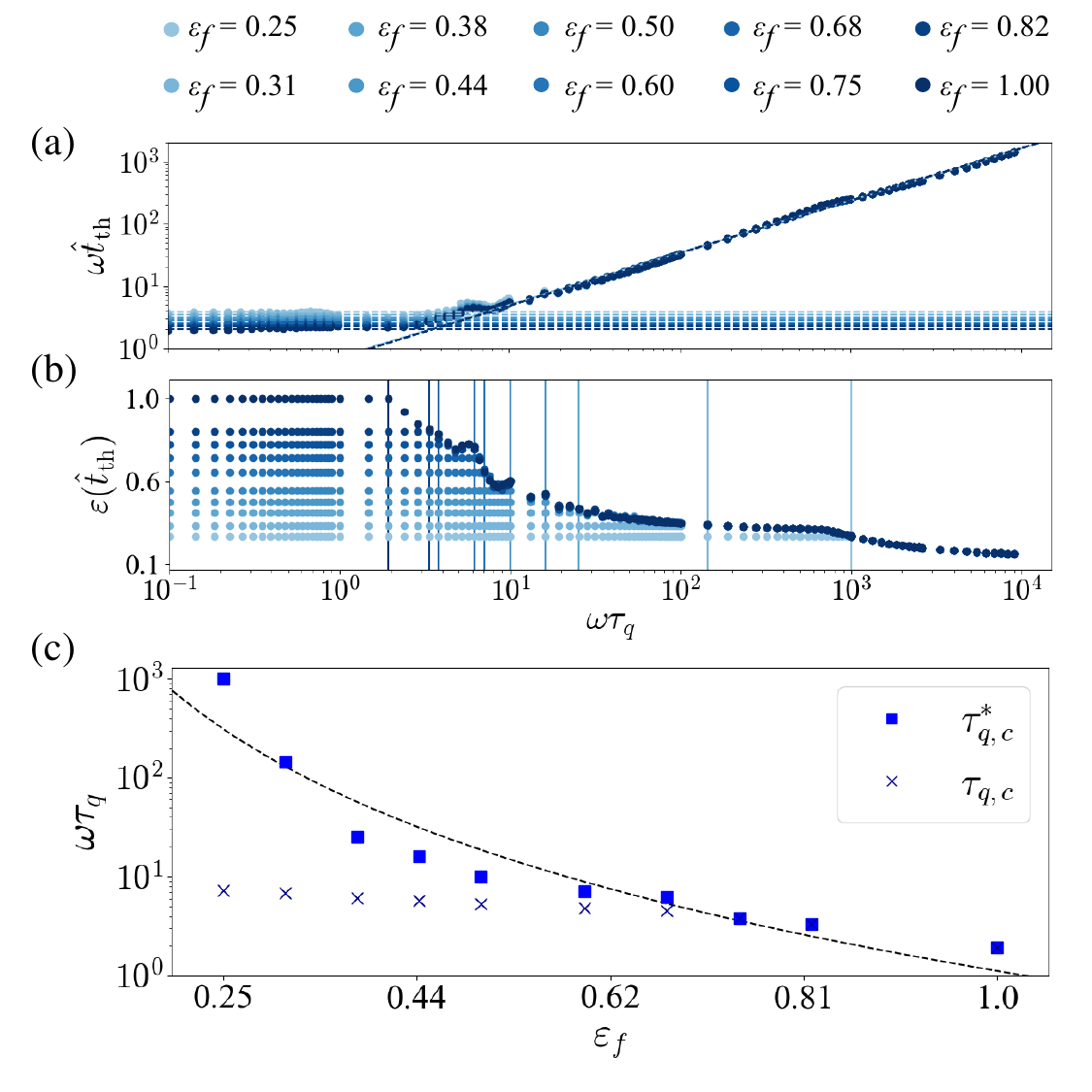}
	\caption{(a)-(b) Scaling of (a) $\hat{t}_{\mathrm{th}}$ and (b) $\varepsilon(\hat{t}_{\mathrm{th}})$ of the open DLM as a function of $\tau_{q}$ for different $\varepsilon_{f}$. The dashed lines in (a) correspond to the best-fit lines of $\hat{t}_{\mathrm{th}}$ for fast and slow quenches, with their intersections marking $\tau_{q, c}$. Meanwhile, the dashed lines in (b) denote $\tau_{q, c}^{*}$. (c) Borders of the delayed KZM regime as a function of $\varepsilon_{f}$. The dashed line corresponds to the best-fit line of $\tau_{q, c}^{*}$. The remaining parameters are $\kappa = J = 0.1\omega$, and $|a|^{2}_{\mathrm{th}}/M = 2$, with $M = 500$ sites.}
	\label{fig: criticality_delayed_kzm}
\end{figure}

Given that $\varepsilon_{f}$ do not alter the behavior of the decay rate of $|a|^{2}$, it is natural to ask whether varying $\varepsilon_{f}$ has any significant effect as well on the scaling of $\hat{t}_{\mathrm{th}}$ and $\varepsilon(\hat{t}_{\mathrm{th}})$, and on the signatures of the delayed KZM. We answer the first question in Figs.~\ref{fig: criticality_delayed_kzm}(a) and \ref{fig: criticality_delayed_kzm}(b), where we show the scaling of $\hat{t}_{\mathrm{th}}$ and $\varepsilon(\hat{t}_{\mathrm{th}})$, respectively, with $\tau_{q}$ for different values of $\varepsilon_{f}$. We can see that varying $\varepsilon_{f}$ does not significantly change the scaling of the KZM quantities considered. However, the $\tau_{q, c}^{*}$, shown as solid lines in Fig.~\ref{fig: criticality_delayed_kzm}(b), increases significantly as we decrease $\varepsilon_{f}$. This modification on $\tau_{q, c}^{*}$ becomes more apparent in Fig.~\ref{fig: criticality_delayed_kzm}(c), where we show the scaling of $\tau_{q, c}$ and $\tau_{q, c}^{*}$ as a function of $\varepsilon_{f}$. Notice that both quantities are inversely proportional to $\varepsilon_{f}$, with $\tau_{q, c}^{*}$ dropping faster than $\tau_{q, c}$ as $\varepsilon \rightarrow \infty$. As a result, the delayed KZM regime vanishes for large $\varepsilon_{f}$, highlighting that its signatures become more apparent for strongly dissipative systems quenched near criticality. We finally note that $\tau_{q, c}^{*}$ follows a power-law scaling as evidenced by the power-law fit curve shown in Fig.~\ref{fig: criticality_delayed_kzm}(c). In particular, since $\tau_{q, c}^{*}$ becomes the true critical quench time separating the breakdown and validity of the KZM at large $\varepsilon_{f}$, we expect that it should follow the power-law scaling \cite{zeng_universal_2023}
\begin{equation}
\label{eq: scaling_tau_qc}
\tau_{q, c}^{*} \propto \varepsilon_{f}^{-\left(vz + 1 \right)},
\end{equation}
which we show to be the case in Appendix \ref{sec: kzm_exponents}.

So far, we have shown that the presence of the delayed KZM leads to a significant deviation between the true freeze-out time, $\hat{t}$, and the transition time, $\hat{t}_{\mathrm{th}}$. Given that the delayed KZM becomes more prominent near criticality at strong dissipation, we now address in the next section how the threshold-based criterion for determining $\hat{t}_{\mathrm{th}}$ contributes to the deviation and whether a more accurate method can be used to measure $\hat{t}$.

\section{Transition Time Measurement}\label{sec: transition_time}

The threshold value used to determine the transition time plays a role in the delay between $\hat{t}$ and $\hat{t}_{\mathrm{th}}$. In particular, we can expect a longer delay for larger $|a|^{2}_{\mathrm{th}}$ since the system's order parameter has to reach a larger threshold value before being detected. This intuition prompts the question of whether decreasing the threshold value has any effect on the scaling of $\hat{t}_{\mathrm{th}}$ and $\varepsilon(\hat{t}_{\mathrm{th}})$, and as to whether it can suppress the deviation brought by the delayed KZM, and thus its signatures.

\begin{figure}
\centering
\includegraphics[scale=0.45]{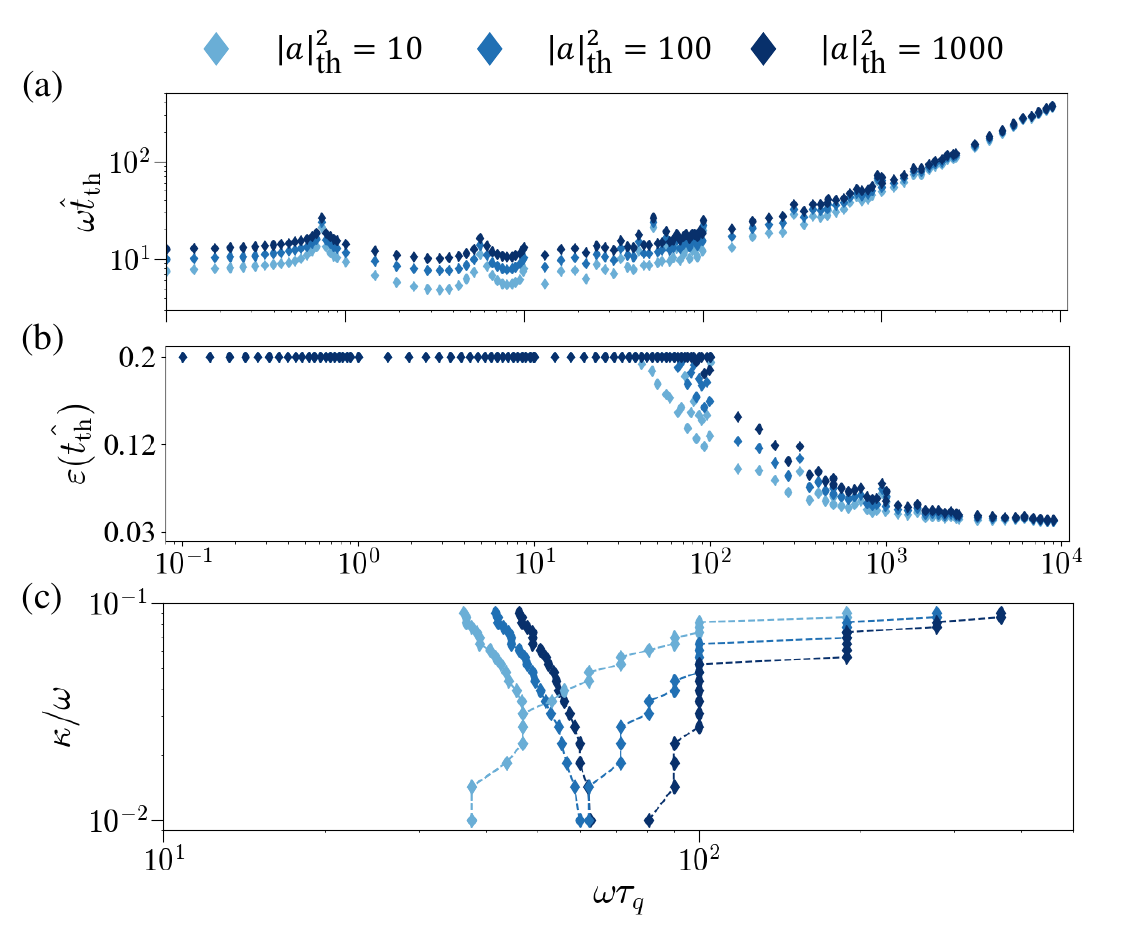}
\caption{(a), (b) Scaling of (a) $\hat{t}_{\mathrm{th}}$ and (b) $\varepsilon(\hat{t}_{\mathrm{th}})$ of the open DM as a function of $\tau_{q}$ for different values of $|a|^{2}_{\mathrm{th}}$. (c) Boundary of the delayed KZM regime as a function of $\kappa$ for different $|a|^{2}$. The remaining parameters are set to $\varepsilon_{f} = 0.2$ and $\kappa = 0.1\omega$.}
\label{fig: threshold_effects}
\end{figure}

We answer the first question in Figs.~\ref{fig: threshold_effects}(a) and \ref{fig: threshold_effects}(b), where we present the scaling of $\hat{t}_{\mathrm{th}}$ and $\varepsilon(\hat{t}_{\mathrm{th}})$, respectively. For this part, while we only consider the open DM, the results here should apply to the COS and the open DLM as well. We can observe that the scaling of $\hat{t}_{\mathrm{th}}$ and $\varepsilon(\hat{t}_{\mathrm{th}})$ do not significantly change as we increase the threshold value. In particular, while the $\hat{t}_{\mathrm{th}}$ is only shifted by a constant value as $|a|^{2}_{\mathrm{th}}$ increases, both KZM quantities considered eventually collapse in a single scaling as $\tau_{q} \rightarrow \infty$. As for the boundaries of the delayed KZM regime, we can observe in Fig.~\ref{fig: threshold_effects}(c) that the gap between $\tau_{q, c}$ and $\tau_{q, c}^{*}$ widens as we increase $|a|^{2}_{\mathrm{th}}$, implying that the delayed KZM becomes more prominent at large $|a|^{2}_{\mathrm{th}}$.

We can understand the widening of the delayed KZM regime for large $|a|^{2}_{\mathrm{th}}$ by noting that in an ideal setup where $\hat{t}$ can be accurately identified, the gap between $\tau_{q, c}$ and $\tau_{q, c}^{*}$ vanishes, and thus following the prediction in Ref.~\cite{zeng_universal_2023}, $\tau_{q, c} = \tau_{q, c}^{*} = \hat{t} / \varepsilon_{f}$. Since for any threshold-based criterion, $\tau_{q, c}^{*} = \hat{t}_{\mathrm{th}} /\varepsilon_{f}$ and $\tau_{q, c} \neq \tau_{q, c}^{*}$ for large $\kappa$ and small $\varepsilon_{f}$, then 
\begin{equation}
\tau_{q, c}^{*} - \tau_{q, c} = \frac{1}{\varepsilon_{f}} \left( \hat{t}_{\mathrm{th}} - \hat{t} \right)
\end{equation}
Let us assume that within the time interval $[\hat{t}, \epsilon_{f} \tau_{q}]$, the total occupation is exponentially growing such that $|a|^{2} \propto  \exp(\gamma_{g}t)$, where $\gamma_{g}$ is the growth rate of the total occupation number. This assumption is supported by Fig.~\ref{fig: open_dm_and_dlm}(c), where the $|a|^{2}$ of the open DM exponentially grows from the minimum value to its saturation value. With this assumption, we can infer that $\hat{t}_{\mathrm{th}} \propto \ln|a|^{2}_{\mathrm{th}} / \gamma_{g}$ and $\hat{t} \propto  \ln|a|^{2}_{\mathrm{min}} / \gamma_{g}$, where $|a|^{2}_{\mathrm{min}}$ is the minimum value of the total occupation number. Thus,
\begin{equation}
\label{eq: delayed_kzm_gap_scaling}
\tau_{q, c}^{*} - \tau_{q, c} \propto \frac{1}{\varepsilon_{f} \gamma_{g}} \left( \ln|a|^{2}_{\mathrm{th}} - \ln|a|^{2}_{\mathrm{min}}  \right).
\end{equation}
which implies that we can suppress the signatures of the delayed KZM by setting $|a|^{2}_{\mathrm{th}}$ close to $|a|^{2}_{\mathrm{min}}$.

\begin{figure}
	\centering
	\includegraphics[scale=0.45]{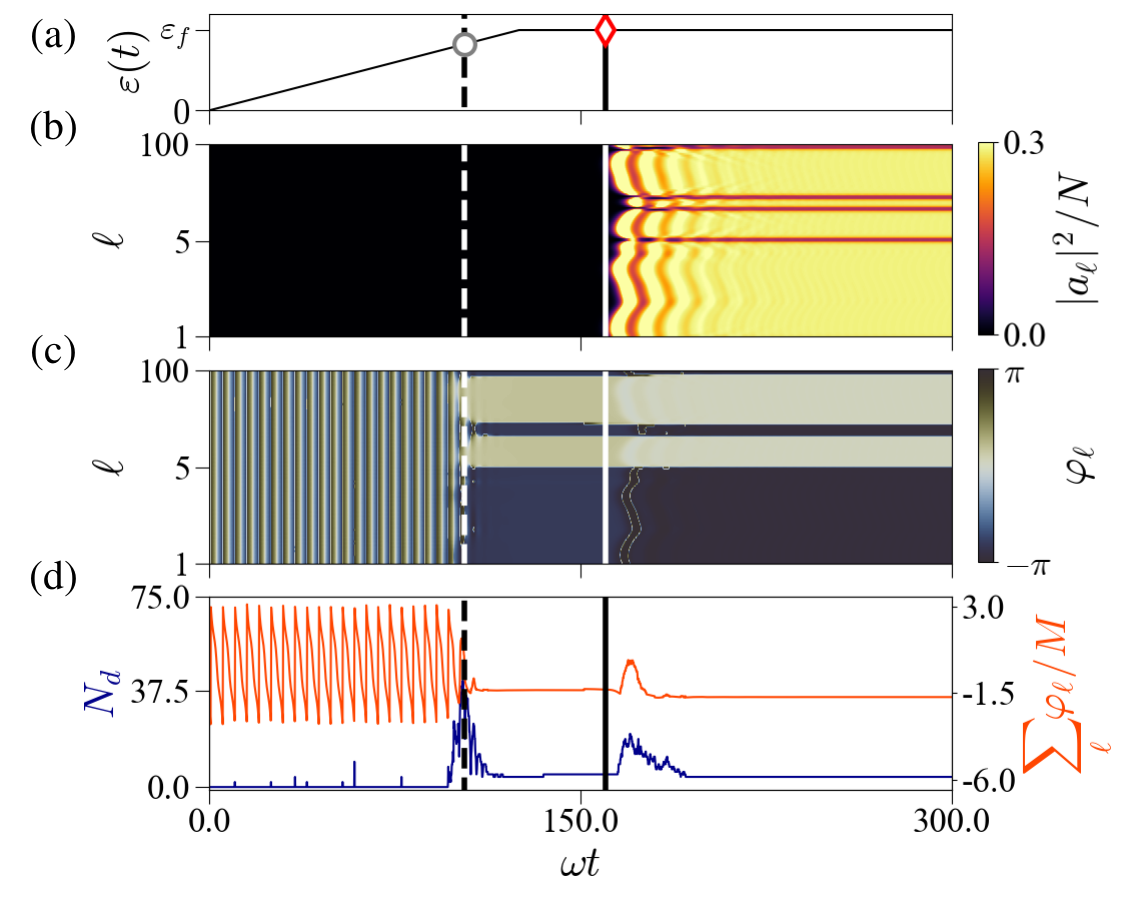}
	\caption{(a) Ramp protocol for the quench time within the $\tau_{q}$-regime of the delayed KZM for the open DLM. The circle point marks $\varepsilon(\hat{t})$, while the diamond point denotes $\varepsilon(\hat{t}_{\mathrm{th}})$. (b), (c) Spatiotemporal dynamics of (b) the occupation number $|a_{\ell}|^{2}$ and (c) the phase of the $\hat{a}_{\ell}$-mode, $\varphi_{\ell}$. (d) Exemplary dynamics of the defect number and the site-averaged $\varphi_{\ell}$. The vertical dashed lines correspond to $\omega \hat{t}$, while the solid lines represent $\omega \hat{t}_{\mathrm{th}}$. The remaining parameters are set to $\varepsilon_{f} = 1.25$, $\kappa = J = 0.1 \omega$, and $|a|^{2}_{\mathrm{th}} = 0.05|a|^{2}_{s}$, where $|a|^{2}_{s}$ is the steady state of $|a|^{2}$.}
	\label{fig: transition_time_measurement}
\end{figure}

Now, determining an optimal threshold value that suppresses the signatures of the delayed KZM may be difficult to achieve as it requires prior knowledge of $|a|^{2}_{\mathrm{min}}$ for arbitrary $\tau_{q}$. This problem motivates the question of whether an alternative method can be used to infer $\hat{t}$ without relying on any threshold-based criterion. As we have hinted in the dynamics of the phase of the $\hat{a}$ mode of the open DM shown in Fig.~\ref{fig: open_dm_and_dlm}(d), we can do this by choosing an appropriate order parameter that rapidly reaches its steady state upon the system entering a phase transition. In the case of the open DM, this order parameter corresponds to the boson mode's phase, $\varphi$. We demonstrate this method further in Figs.~\ref{fig: transition_time_measurement}(b) and \ref{fig: transition_time_measurement}(c), where we show that for nonzero dimensional systems, like the open DLM, we can use the phase information of the $\hat{a}_{\ell}$ modes to extract $\hat{t}$. As presented in Fig.~\ref{fig: transition_time_measurement}(d), we can do this by determining the time at which either the defect number, $N_{d}$, or the site-averaged phase begins to saturate. In the COS level, the inferred $\hat{t}$ for this method would be equivalent to the moment the system picks a new global minimum it would fall onto, signaling phase transition. Thus, we expect that if the system's phase information is available in an experimental setup, such as in Ref.~\cite{kesler_observation_2021}, then that can serve as a more sensitive tool for detecting phase transitions compared to threshold-based order parameters that depend on the mode occupations.

\section{Summary and Discussion}\label{sec: conclusion}

In this paper, we extend the Kibble-Zurek mechanism to open systems under a finite quench and report an intermediate regime separating the breakdown and validity of the KZM at fast and slow quench timescales, respectively. This regime manifests as a continuation of the transition time's KZM power-law scaling at $\tau_{q}$, where the system appears to relax after the quench has terminated. As we have shown using a coupled oscillator system, this phenomenon results from the system's relaxation towards the global minimum of its potential due to dissipation. This mechanism effectively hides the system's crossover to the adiabatic regime, only to be revealed once the system reaches the arbitrary threshold of the order parameter.

Using the open DM and the open DLM, we have also demonstrated that the delayed KZM is a generic feature of open systems under finite quenches that can be mapped onto a coupled oscillator system. Furthermore, we have shown that the signatures of the delayed KZM, specifically the size of the quench interval where the delayed KZM regime is observed, become more prominent for small values of $\varepsilon_{f}$, highlighting the dissipative and near-critical nature of this phenomenon. We have discussed the implications of the delayed KZM in the context of the threshold-based criterion typically used in experiments to measure the transition time and proposed an alternative method to measure $\hat{t}$. Our proposed method only relies on the spatialtemporal information of an appropriate order parameter, such as the defect number and phase information of the system's bosonic modes, thus providing a more sensitive tool for detecting phase transitions.

Our results extend the notion of the KZM to dissipative systems with finite quench protocols beyond the limits of slow and rapid quenches. It also provides a framework on how the manifestation of the KZM can be altered in experimental protocols, wherein limitations in measuring the true AI crossover become more relevant. Since our results are all in the mean-field level, a natural extension of our paper is to verify whether the delayed KZM would survive in the presence of quantum fluctuations. It would also be interesting to test the signatures of the delayed KZM in the quantum regime of the open DM and the open DLM, and further explore their universality classes beyond the mean-field level. These extensions can be readily done in multiple platforms, including, but not limited to, cavity-QED setups \cite{scigliuzzo_controlling_2022, white_cavity_2019, zhu_interplay_2020, klinder_dynamical_2015, baumann_dicke_2010}, nitrogen-vacancy center ensembles \cite{zou_implementation_2014, amsuss_cavity_2011,liu_entanglement_2016,  astner_coherent_2017}, cavity-magnon systems \cite{rameshti_cavity_2022, kim_observation_2024}, and photonic crystals \cite{paternostro_solitons_2009}.

\section*{Acknowledgements}
This work was funded by the UP System Balik PhD Program (OVPAA-BPhD-2021-04) and the DOST-SEI Accelerated Science and Technology Human Resource Development Program.

\appendix

\section{Mean-field Equations of the Considered Systems}\label{sec: mean_field_eom}

To obtain the mean-field equations of the systems considered in the main text, we consider the master equation for the expectation value of an arbitrary operator, $\hat{O}$,
\begin{equation}
\label{eq:master_eq}
    \partial_{t} \left< \hat{O} \right> = i \left< \left[ \frac{\hat{H}}{\hbar}, \hat{O} \right]  +   \mathcal{D}\hat{O} \right>, 
\end{equation}
where $\hat{H}$ is the system's Hamiltonian, and $\mathcal{D}\hat{O} = \sum_{\ell} \kappa_{\ell} \left( 2\hat{L}_{\ell}^{\dagger} \hat{O} \hat{L}_{\ell} - \left\{ \hat{L}^{\dagger}_{\ell} \hat{L}_{\ell}, \hat{O} 
 \right\} \right)$ is the dissipator, with $\hat{L}_{\ell}$ being the jump operators. We will also let $A = \left< \hat{A} \right>$ for notation convenience. Using this master equation, the mean-field equation of the open COS is 
\begin{subequations}
\label{eq: COS_eom}
\begin{equation}
	\partial_{t} a = -i \left[ \omega a + \lambda \left( b + b^{*} \right) \right] - \kappa a,
\end{equation}
\begin{equation}
	\partial_{t} b = -i \left[ \omega_{0} b + \lambda \left( a + a^{*} \right] \right).
\end{equation}
\end{subequations}
As for the open Dicke model, its mean-field equations are
\begin{subequations}
\label{eq:open_dm_eom}
    \begin{equation}
        \partial_{t} a = -i \left( \omega a + \frac{2\lambda}{\sqrt{N}} S^{x} \right) - \kappa a, 
    \end{equation}
	\begin{equation}
		\partial_{t} S^{x} = - \omega_{0}S^{y},
	\end{equation}	    
    \begin{equation}
    		\partial_{t} S^{y} = \omega_{0} S^{x} - \frac{2\lambda}{\sqrt{N}} \left( a^{*} + a \right) S^{z},    
    \end{equation}
    \begin{equation}
 \quad \quad \partial_{t} S^{z} = \frac{2\lambda}{\sqrt{N}}\left( a^{*} + a \right) S^{y}.
    \end{equation}
\end{subequations}
Note that for both the open COS and open DLM, the jump operator is given to be $\hat{L}_{\ell} = \hat{L} = \hat{a}$. Finally, the mean-field equations of the open DLM for a jump operator $\hat{L}_{\ell} = \hat{a}_{\ell}$ is
\begin{subequations}
\label{eq:odlm_eom}
    \begin{equation}
    \partial_{t} a_{\ell} = -i \left[ \omega a_{\ell} + \frac{2\lambda}{\sqrt{N}} S^{x}_{\ell} - J \left( a_{\ell-1} + a_{\ell+1} \right) \right] - \kappa a_{\ell}, 
    \end{equation}
    	\begin{equation}
       \partial_{t} S^{x}_{\ell} = - \omega_{0} S^{y}_{\ell},
    	\end{equation}
    	\begin{equation}
	    \partial_{t} S^{y}_{\ell} = \omega_{0}S^{x}_{\ell} - \frac{2\lambda}{\sqrt{N}} \left( a_{\ell} + a^{*}_{\ell} \right)S^{z}_{\ell},    	
    	\end{equation}
    \begin{equation}
	    \partial_{t} S^{z}_{\ell} = \frac{2\lambda}{\sqrt{N}}\left( a_{\ell} + a^{*}_{\ell} \right)S^{y}.
    \end{equation}
\end{subequations}

\section{KZM exponents of the open Dicke Lattice model}\label{sec: kzm_exponents}

\begin{figure}
\includegraphics[scale=0.475]{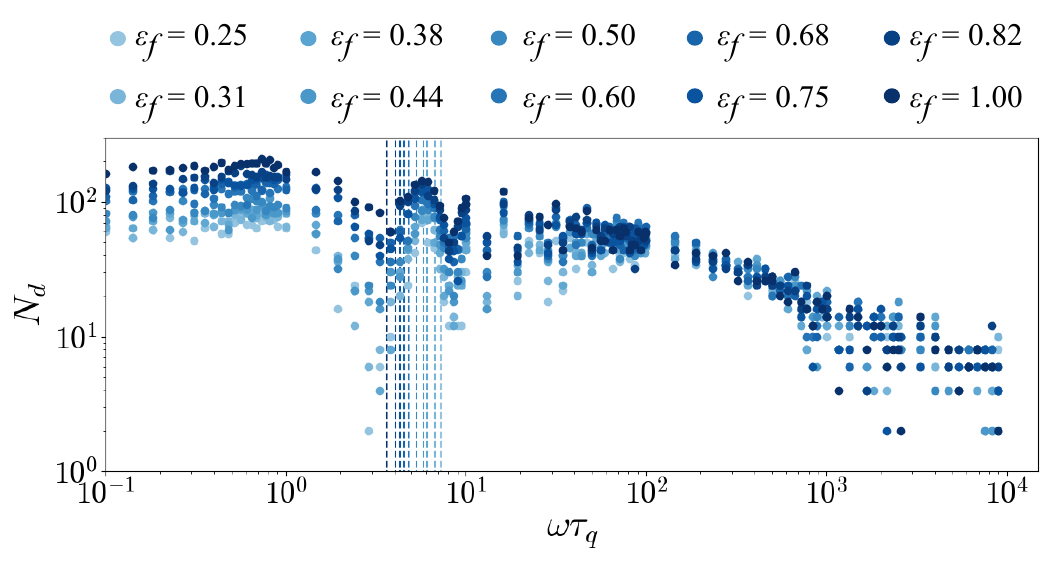}
\caption{Scaling of the defect number, $N_{d}$, of the open DLM as a function of $\tau_{q}$, for different values of $\varepsilon_{f}$. The vertical dashed lines correspond to $\tau_{q, c}$ for a given $\varepsilon_{f}$. The system parameters are set to $\kappa=J = 0.1 \omega$.}
\label{fig: defect_scaling}
\end{figure}

One of the key predictions of the KZM is the power-law scaling of the defect number as a function of $\tau_{q}$ \cite{del_campo_universality_2014},
\begin{equation}
\label{eq: defect_number_scaling}
N_{d}  \propto \tau_{q}^{-\left(D -d \right) \frac{v}{1 + vz} },
\end{equation}
where $D$ and $d$ are the dimensions of the system and the topological defects, respectively. To demonstrate that the open DLM satisfies the predicted KZM scaling for $N_{d}$, we present in Fig.~\ref{fig: defect_scaling} the number of phase slips present in the system for a given $\tau_{q}$ and $\varepsilon_{f}$. We can observe that for large $\tau_{q}$, $N_{d}$ follows a power-law scaling behavior with $\tau_{q}$, emphasizing that the system indeed follows the KZM at slow quenches, which is consistent with the behavior of $\hat{t}$ and $\varepsilon(\hat{t})$ shown in Figs.~\ref{fig: criticality_delayed_kzm}(a) and \ref{fig: criticality_delayed_kzm}(b). Notice, however, that as we approach $\tau_{q, c}$, marked by the vertical dashed lines, $N_{d}$ starts to fluctuate, with the fluctuation becoming more significant as $\varepsilon_{f} \rightarrow 0$. This behavior is akin to the presaturation regime observed for closed systems under finite quench protocols \cite{kou_varying_2023}. As to whether this regime persists in the presence of quantum fluctuation remains an open question. We finally observe the saturation of $N_{d}$ as $\tau_{q} \rightarrow 0$, signifying the breakdown of the KZM for small values of $\tau_{q}$.

Since we have shown that the KZM quantities $\hat{t}_{\mathrm{th}}$, $\varepsilon(\hat{t}_{\mathrm{th}})$, and $N_{d}$ follow the predicted KZM scaling, for completeness, we now estimate the critical exponents of the open DLM from the power-law exponents of the of these quantities. We do this by assuming that $\hat{t}_{\mathrm{th}}$ and $N_{d}$ follows a generic power-law scaling, 
\begin{equation}
\label{eq: kzm_powerlaw_ansatz}
\hat{t}_{\mathrm{th}} \propto \tau_{q}^{\alpha}, \quad N_{d} \propto \tau_{q}^{\beta},
\end{equation}
within the quench time interval $\omega \tau_{q} = \omega \tau_{q, c}$ and $\omega \tau_{q} = 10^{4}$. From these equations, we can infer from Eqs.~\eqref{eq: kzm_predicted_scaling} and Eq.~\eqref{eq: defect_number_scaling} that $\alpha$ and $\beta$ are related to critical exponents $v$ and $z$ by the relations
\begin{equation}
v = \frac{\alpha}{|\beta|}, \quad z = \frac{|\beta|}{1 - \alpha}, \quad vz = \frac{\alpha}{1 - \alpha}.
\end{equation}

\begin{figure}
\includegraphics[scale=0.35]{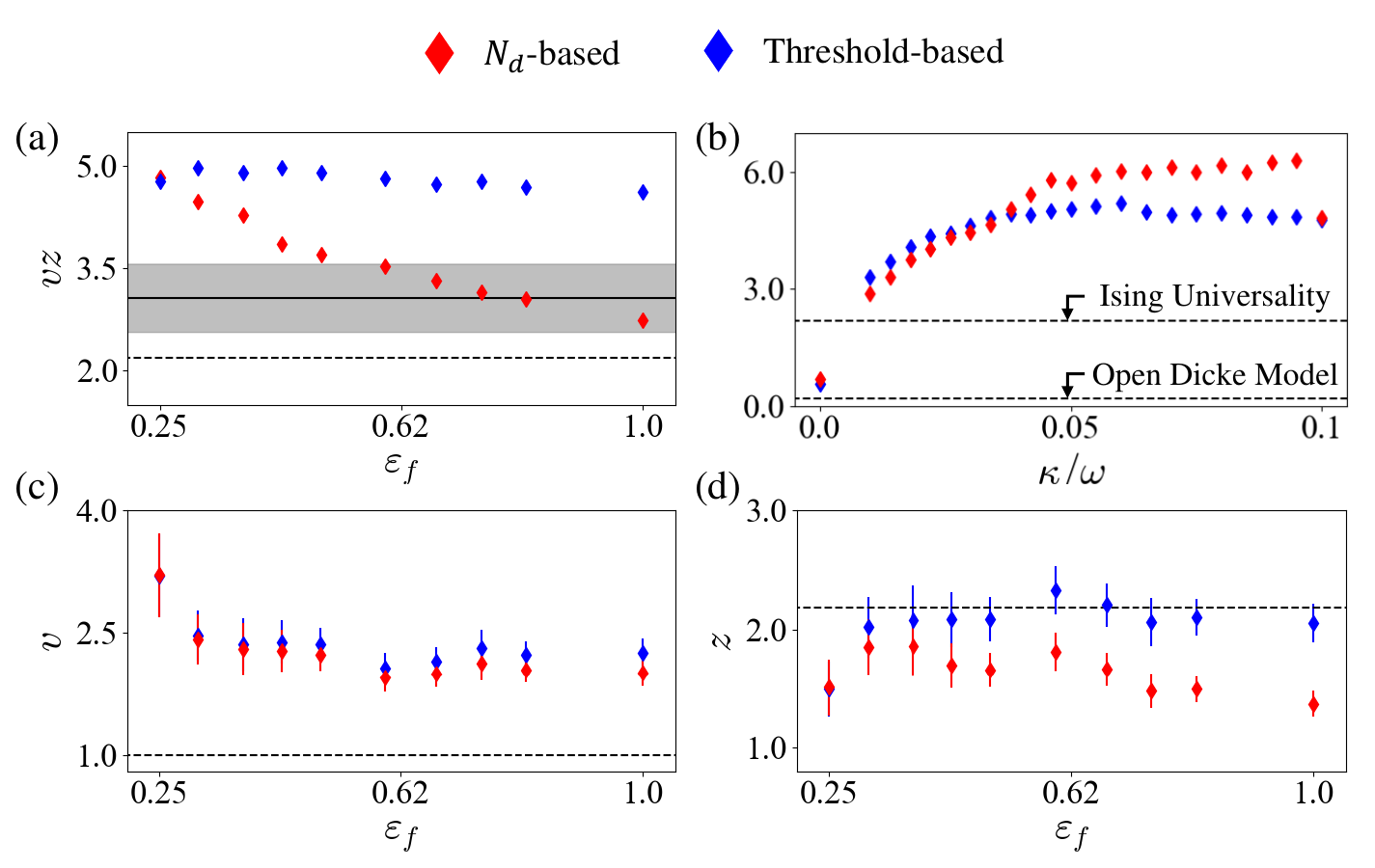}
\caption{(a), (b) Calculated $vz$ from the $N_{d}$ and threshold-based transition time as a function of (a) $\varepsilon_{f}$ for $\kappa = 0.1\omega$ and (b) $\kappa$ for $\varepsilon_{f} = 1.25$. The remaining parameter is $J=0.1\omega$. (c), (d) Estimated critical exponents (c) $v$ and (d) $z$ as a function of $\varepsilon_{f}$. The dashed lines correspond to the critical exponents of the Ising universality class \cite{delfino_integrable_2004, dammann_dynamical_1993}. The remaining parameters are set to $\kappa = J = 0.1\omega$.}
\label{fig: kzm_critical_exponents}
\end{figure}

We present in Figs.~\ref{fig: kzm_critical_exponents}(a) and \ref{fig: kzm_critical_exponents}(b) the estimated $vz$ as a function of $\varepsilon_{f}$ and $\kappa$, respectively, for both the threshold-based transition time and the $\hat{t}$ obtained from the dynamics of the $N_{d}$, as described in Sec.~\ref{sec: transition_time} of the main text. We can see that the threshold-based $vz$ remains relatively constant for all values of $\varepsilon_{f}$, while the defect-based $vz$ appears to converge to the critical exponent of the Ising universality \cite{delfino_integrable_2004, dammann_dynamical_1993}, the universality class of the single Dicke model \cite{kirton_introduction_2019}. We further check whether the two values of $vz$ are consistent with one another by calculating $vz$ as well from the scaling of $\tau_{q, c}^{*}$ with $\varepsilon_{f}$, which is given by Eq.~\eqref{eq: scaling_tau_qc}. We show this in Fig.~\ref{fig: kzm_critical_exponents}(a) as a solid line, with the grey regions corresponding to the uncertainty due to fitting errors. Notice that the value of $vz$ from the defect-based method is consistent with the one obtained from $\tau_{q, c}^{*}$ for large values of $\varepsilon_{f}$, while it becomes more consistent with the threshold-based $vz$ for small $\varepsilon_{f}$. With this picture, the threshold-based $vz$ and $\tau_{q, c}^{*}$ can be interpreted as the upper and lower bounds for the uncertainty of the open DLM's critical exponents in the mean-field level, respectively.

As for the behavior of the threshold-based and defect-based $vz$ as a function of $\kappa$, we can see in Fig.~\ref{fig: kzm_critical_exponents}(b) that both values of $vz$ decrease as $\kappa \rightarrow 0$. In particular, the $vz$ for both cases approaches the experimental value of $vz$ for the open Dicke model for $\kappa = 1.0\omega$ \cite{klinder_dynamical_2015}. This result implies that the system's dissipation modifies the critical exponents of the system, which is consistent with the predictions in Refs.~\cite{rossini_dynamic_2020} and \cite{hedvall_dynamics_2017}. Without any specific analytical prediction on how $\kappa$ modifies the effective critical exponent of the system, we cannot assign a universality class for the open DLM that may apply to any arbitrary dissipation strength.

Given this limitation, we restrict the calculation of the critical exponents for $\kappa = 0.1\omega$. We show in Figs.~\ref{fig: kzm_critical_exponents}(c) and \ref{fig: kzm_critical_exponents}(d) the values of $v$ and $z$, respectively, for both the defect-based and threshold-based methods. We can observe that the value of $v$ for both cases has a large deviation from the static critical exponent of the Ising universality, which is $v = 1$ \cite{delfino_integrable_2004}. Meanwhile, the value of $z$ for the threshold-based criterion converges to $z \sim 2.183$, which is the dynamic critical exponent of the Ising universality class \cite{dammann_dynamical_1993}. As we previously mentioned, we can attribute the deviations of $v$ and $z$ to the dissipation-induced modification of the critical exponents. The accumulated errors on the scaling exponents of $N_{d}$, $\hat{t}_{\mathrm{th}}$, and $\hat{t}_{N_{d}}$ due to the fitting errors may also amplify the deviations of the critical exponents from their expected values. Determining which case has a more significant effect on the values of $v$ and $z$ requires understanding the dynamics of the open DLM beyond the mean-field level.

\bibliography{reference.bib}
\bibliographystyle{apsrev4-2}

\end{document}